\documentstyle[aps,preprint,epsfig,floats]{revtex}

\tightenlines

\begin{document}

\title{The High $E_T$ Drop of $J/\psi$ to Drell-Yan Ratio\\ 
from the Statistical $c\bar{c}$ Coalescence  Model}

\author{
A.P. Kostyuk$^{a,b}$,
%\footnote{E--mail: goren@th.physik.uni-frankfurt.de},
M.I. Gorenstein$^{b}$,
%\footnote{E--mail: kostyuk@th.physik.uni-frankfurt.de},
H. St\"ocker$^{a}$,
%\footnote{E--mail: stoecker@th.physik.uni-frankfurt.de}
and
W. Greiner$^{a}$
%\footnote{E--mail: greiner@th.physik.uni-frankfurt.de}
}

\address
{$^a$ Institut f\"ur Theoretische Physik, Universit\"at  Frankfurt,
Germany}

\address{$^b$ Bogolyubov Institute for Theoretical Physics,
Kyiv, Ukraine}

\date{\today}
\maketitle

\begin{abstract}
The dependence of the $J/\psi$ yield on the
transverse energy $E_T$ in heavy ion collisions is considered within
the statistical $c\bar{c}$ coalescence model.  The model fits the NA50 
data for Pb+Pb collisions at the CERN SPS even in  
the high-$E_T$ region ($E_T\agt 100$~GeV). Here  
$E_T$-fluctuations and $E_T$-losses in the dimuon event sample 
naturally create the celebrated drop in the $J/\psi$ to Drell-Yan
ratio.
\end{abstract}


Measurements of the $J/\psi$ suppression pattern in heavy ion
collisions had been proposed as a diagnostic tool for the 
deconfinement of strongly interacting
matter at the early stage of the reaction
\cite{MS} (see also \cite{Satz} for a modern review). However, it was found
that the $J/\psi$ suppression in proton-nucleus and
nucleus-nucleus collisions with light projectiles (up to S+U) can
be explained by inelastic collisions with the nucleons of the incident
nuclei (the so-called `normal nuclear suppression') \cite{NA38}.
In contrast, the NA50 experiments with a heavy projectile and a
heavy target (Pb+Pb) revealed {\it `anomalous'} $J/\psi$
suppression \cite{anomalous,threshold,evidence}, i.e. an
essentially stronger drop of the $J/\psi$ over Drell-Yan ratio $R$
than it could be expected from a simple extrapolation of the light
projectile data. The dependence of  $R$
on the neutral transverse energy $E_T$, which is used as
an estimator of the collision centrality, shows a two step
behavior. The {\it onset} \cite{threshold} of the anomalous
$J/\psi$ suppression (deviation of the data downwards from the
`normal nuclear suppression' curve) supposedly takes place at $E_T \approx
35$~GeV. The second drop has been noted at $E_T \approx
100$~GeV. The initial interpretation \cite{evidence} attributed
these two drops to the successive melting of the charmonium $\chi_c$
states (which contribute about 45\% to the measured $J/\psi$ number) and
primary $J/\psi$'s in the quark-gluon plasma.

Other interpretations are, however, possible. Because of
relatively high statistical errors of the data at low $E_T$, the
behavior around $E_T \approx 40$~GeV can be equally well
interpreted as smooth decreasing of the measured ratio $R(E_T)$ due to
charmonium suppression by comoving hadrons\footnote{Later it was
found that even the sweeping nucleons of the colliding nuclei
might, under certain assumptions, produce the observed suppression
\cite{Qiu}.} \cite{comover}. Discussing the second drop of the
$J/\psi$ to Drell-Yan ratio, one has to take into account
\cite{nonsaturation} that the neutral transverse energy $E_T$
estimates different quantities in the regions below and above $E_T
\approx 100$~GeV. At $E_T \alt 100$~GeV, it is proportional with
good accuracy to the number of nucleon participants $N_p$ and
provides a reliable measure of the collision centrality. The 
large $E_T$ `tail', however, is produced mainly due to
fluctuations of the transverse energy at almost fixed (nearly
maximal) centrality. The influence of these fluctuations on the
$J/\psi$ suppression can partially explain the second drop. Better
agreement of the co-mover model with the data in the region  $E_T
\agt 100$ is achieved when a loss of transverse energy in the
dimuon data sample with respect to the minimum bias one is also taken
into account \cite{E_T_loss}.

The $E_T$ fluctuations appear to be important in the deconfinement
scenario as well. It was found \cite{Mai}
that while the first drop can be explained
by melting of $\chi_c$'s, the onset of
primary $J/\psi$ melting does not produce any visible structure on the curve.
The second drop appears due to the fluctuations of $E_T$.

All mentioned approaches \cite{MS,Qiu,comover,nonsaturation,E_T_loss,Mai},
despite of their differences, have one common feature: charmonia are
assumed to be created {\it exclusively} at the initial stage of the
reaction in primary nucleon collisions. Any subsequent interactions
(with sweeping nucleons, co-moving hadrons or quark-gluon medium) may only
destroy them. Despite of rather successful agreement with the $J/\psi$ 
data, such a scenario seems to be in trouble explaining the $\psi'$ yield.
The recent lattice simulations \cite{Karsch} suggest that the temperature of 
$\psi'$ dissociation $T_d(\psi')$ lies far below the deconfinement point 
$T_c$: $T_d(\psi') \approx 0.1$--$0.2 T_c$ \cite{Satz01}. Therefore, not 
only the quark-gluon plasma, but also a hadronic co-mover medium should 
completely eliminate $\psi'$ charmonia in central Pb+Pb collisions at SPS. 
However, the experiment revealed a sizable $\psi'$ yield (see, for instance, 
\cite{Bordalo,Jaipur}). It was observed \cite{Shuryak} that
$\psi'$ to $J/\psi$ ratio decreases with centrality only in peripheral
lead-lead collisions, but remains approximately constant and equal to 
its thermal value at sufficiently large number of participants 
$N_p \ge 100$.

A natural way to explain the $\psi'$ yield in nucleus-nucleus collisions 
without running into contradiction with the lattice data is to assume 
that a charmed quark and an antiquark that 
have been created at the initial (`hard') stage of the reaction can 
coalesce to form a hidden charm meson at a later stage,
close to the chemical freeze-out point.
This possibility has been considered in the literature
\cite{GG,Br1,Go:00,Le:00,Ka:00,Ra:00}
within various models.

The statistical $c\bar{c}$ coalescence model of $J/\psi$ production 
suggested in \cite{Br1,Go:00} will  be considered in the present paper. 
The charmed quark-antiquark ($c\bar{c}$) pairs are assumed to be
created at the initial stage of the nucleus-nucleus reaction in
hard parton collisions. Creation of $c\bar{c}$ pairs after the
hard initial stage as well as their possible annihilation is
neglected. Therefore, the number of charmed quark-antiquark pairs
remains approximately unchanged during the subsequent evolution.
They are then distributed over open and hidden charm
particles at the final stage of the reaction in accordance with laws of
statistical mechanics. It is postulated that the deconfined medium prevents
formation of charmonia at early stage of the reaction. In fact, it was 
found \cite{Satz01a} that $J/\psi$ becomes unbound slightly above the
deconfinement point $T_d(J/\psi) \approx 1.1 T_c$. Therefore, all charmonia 
observed in central Pb+Pb collisions supposed to be created at the final
stage of the reaction.  

It was shown in our previous paper \cite{Ko:01} that the
statistical $c\bar{c}$ coalescence model is in excellent agreement with the
NA50 data for (semi)central ($N_p \agt 100$) Pb+Pb collisions.
However, the analysis of \cite{Ko:01} was restricted to the
region $E_T \alt 100$~GeV, where $E_T$ estimates the collision centrality. 
Fluctuations of
$E_T$ were not taken into account. The aim of the present paper is
to extend the applicability domain of the model to $E_T \agt
100$~GeV and describe the second drop of the anomalous $J/\psi$
suppression. Two effects appear to be important in this region:
\begin{itemize}
\item Influence of fluctuations of thermodynamic parameters of
the system on the $J/\psi$ production.
\item A loss of neutral transverse energy, $E_T$, in the event
sample collected with the dimuon trigger relative to the minimum
bias events.
\end{itemize}

As far as  charmed (anti)quarks are created exclusively at the
initial `hard' stage of the reaction in collisions
of primary nucleons,
the average number of $c\bar{c}$
pairs $\langle c\bar{c} \rangle_{AB(b)}$ created in a nucleus-nucleus
(A+B) collision at fixed impact parameter $b$ is proportional to
the number of primary nucleon-nucleon collisions and can be
found in Glauber's approach:
\begin{equation}
\langle c\bar{c} \rangle_{AB(b)} = A B \sigma^{NN}_{c\bar{c}} T_{AB}(b).
\end{equation}
Here $T_{AB}(b)$ is the nuclear overlap function calculated assuming
a Woods-Saxon distribution of nucleons in nuclei \cite{tables}
(see e.g. Appendix of \cite{Ko:01}).
$\sigma^{NN}_{c\bar{c}}$ is the production cross section of a $c\bar{c}$
pair in a nucleon-nucleon collision. There are indirect experimental indications
\cite{NA50open} and a phenomenological model \cite{hf_enh}
suggesting  an enhancement of the open charm in heavy ion collisions
with respect to a direct extrapolation of nucleon-nucleon data.
Therefore, $\sigma^{NN}_{c\bar{c}}$ in nucleus-nucleus collisions is not necessarily
equal to its value measured in a proton-proton experiment.
We consider $\sigma^{NN}_{c\bar{c}}$ as
a free parameter. Its value is fixed by fitting the NA50 data.

The primary nucleon-nucleon collisions in a nucleus-nucleus
reaction are independent and the probability to produce a
$c\bar{c}$ pair in a single collision is small. Therefore,
event-by-event fluctuations of the number of $c\bar{c}$ pairs in
nucleus-nucleus collisions at fixed impact parameter $b$ are
Poisson distributed. Assuming exact conservation of this number
one obtains the following statistical coalescence model result for
the average number of produced $J/\psi$'s per A+B collision
\cite{Go:00}:
\begin{equation}
\label{Jpsi}
\langle J/\psi \rangle_{AB(b)}^{(T,\mu_B)} =
\langle c\bar{c} \rangle_{AB(b)}
\left[ 1 + \langle c\bar{c} \rangle_{AB(b)} \right]
\frac{N_{J/\psi}^{tot}(T,\mu_B)}{(N_O(T,\mu_B)/2)^2} + \dots,
\end{equation}
Where $T$ and $\mu_B$ are, respectively, the system temperature and
baryonic chemical potential. The dots in (\ref{Jpsi}) state for
higher order terms with respect to the ratio 
$\frac{N_{J/\psi}^{tot}(T,\mu_B)}{(N_O(T,\mu_B)/2)^2}$, which can
be safely neglected at $N_p > 100$. Here  $N_O$ is the total open charm
multiplicity calculated within the grand canonical ensemble of the
equilibrium hadron gas model:
\begin{equation}
\label{N_O}
N_O = V \! \! \! \! \sum_{j=D,\bar{D},D^*,\bar{D}^*,\dots} 
\! \! \! \! n_j(T,\mu_B) \equiv
V n_O(T,\mu_B).
\end{equation}
The sum in (\ref{N_O}) runs over all known (anti)charmed particle
species \cite{pdg}.
The total $J/\psi$ multiplicity $N_{J/\psi}^{tot}(T,\mu_B)$
includes the contribution of
excited charmonium states decaying into $J/\psi$:
\begin{equation}
\label{dec}
N_{J/\psi}^{tot}= V \! \! \! \! \sum_{j=J/\psi,\chi_1,\chi_2,\psi'} \! \! \! \!
R(j\rightarrow J/\psi) n_j(T,\mu_B)  \equiv V n_{J/\psi}^{tot}(T,\mu_B).
\end{equation}
Here $R(j \rightarrow J/\psi)$  is the decay branching ratio of the
charmonium $j$ into $J/\psi$:
$R(J/\psi \rightarrow J/\psi) \equiv 1$,
$R(\chi_1 \rightarrow J/\psi)\approx 0.27$,
$R(\chi_2 \rightarrow J/\psi)\approx 0.14$
and $R(\psi^{\prime} \rightarrow J/\psi)\approx 0.54$.
The volume parameter $V$ is fixed by the condition of the baryon
number conservation:
\begin{equation}
\label{bnc}
N_p(b) = V \! \! \! \! \sum_{j=N,\bar{N},\Delta,\bar{\Delta},\dots} \! \! \! \!
b_j n_j(T,\mu_B) \equiv V n_B(T,\mu_B) \ .
\end{equation}
Here $b_j$ is the baryon number of particle type $j$ and
the sum in (\ref{bnc})
runs over all known (anti)baryon species \cite{pdg}.
The number $N_p(b)$ of nucleon participants at fixed impact parameter $b$
is found in Glauber's approach (see e.g. Appendix of \cite{Ko:01})\footnote{In
what follows we neglect fluctuations of the number of
nucleon participants at fixed impact parameter.}.

The particle densities in (\ref{N_O}--\ref{bnc}) are found from:
\begin{equation}\label{Nj}
n_j(T,\mu_B) \ = \  \frac{d_j}{2\pi^2} \int_0^{\infty}k^2dk 
\left[\exp\left(\frac{\sqrt{m_j^2+k^2} - \mu_j}{T}\right) \pm
 1 \right]^{-1} \ ,
\end{equation}
where $m_j$ and $d_j$ are, respectively, the mass and
degeneracy factor of particle type $j$;
$\mu_j=b_j\mu_B+s_j\mu_S+c_j\mu_C$ is the chemical potential
with $s_j$ and $c_j$ being, respectively, the
strangeness and charm of particle type $j$. The strange and charm
chemical potentials  $\mu_S$ and $\mu_C$ are fixed by the requirement of
zero net strangeness and charm in the system.

As was already discussed above, the analysis \cite{Satz01} of the recent
lattice data \cite{Karsch} has shown that $\psi'$ cannot exist in 
hot hadronic medium. To explain the observed $\psi'$ multiplicity one
has to assume that charmonia are formed close to the chemical freeze-out 
point. (For (semi)central Pb+Pb collisions at SPS the chemical freeze-out
(almost) coincides with the hadronization.)  The thermodynamical parameters
$T$ and $\mu_B$ at chemical freeze-out can be
estimated by fitting  the  hadron gas model to the hadron yield
data in Pb+Pb collisions at SPS \cite{Becattini,GY,Br2}. This
procedure, however, provides only {\it average}
(in a sense that will be defined later) values $\overline{T}$
and $\overline{\mu}_B$. In fact, the thermodynamic parameters are
not the same for all events, but rather they are subject to fluctuations of
dynamical origin.

Only a fraction of the initial energy of the incident nuclei
is converted to {\it thermal} energy  $E_{th}$ 
(including the rest masses) of the produced
hadron gas. The rest is redistributed to kinetic energy of the
collective flow. The contribution of each nucleon participant to
$E_{th}$ is not fixed. It fluctuates in accord with a hitherto 
unknown dynamical mechanism.

Suppose one is able to measure both $N_p$ and $E_{th}$. Then,
events can be grouped in bins with different thermal energy
per participant, $E_{th}/N_p$. Different thermal energies per
net baryon number cannot be obtained at the same values of the
temperature and baryonic chemical potential of the hadron gas.
Therefore, fitting  the  hadron gas model to the hadron yield
data in each bin, one would obtain different values of $T$
and $\mu_B$. In reality, however, such a selection of events is
not feasible, therefore no experimental information about
fluctuations of $T$ and $\mu_B$ is available.
The procedure of \cite{Becattini,GY,Br2} has
to mix events with different $E_{th}/N_p$ and estimates {\it average}
values $\overline{T}$ and $\overline{\mu}_B$.
One can, however, make a reasonable assumption about the
fluctuations.  It was found that at different initial energies of colliding
nuclei, the produced hadron gas chemically freezes-out at constant total
thermal energy per hadron \cite{f-o_criterion}:
\begin{equation}
\label{f-o_crit}
E_{th}(T,\mu_B)/N_{h}(T,\mu_B) = \mbox{const} \approx 1 \mbox{ GeV}.
\end{equation}
Different initial energies result in different thermal energy per
participant. Therefore, it is natural to assume that the same criterion
is valid when $E_{th}/N_p$ fluctuates at fixed initial energy.

At given $T$ and $\mu_B$, the thermal energy
of the hadron gas can be found from
\begin{equation}
\label{Eth}
E_{th}(T,\mu_B) = V \sum_{j} \varepsilon_j(T,\mu_j),
\end{equation}
where $\varepsilon_j(T,\mu_B)$ is the contribution of the hadron
species $j$ to the thermal energy density:
\begin{equation}
\label{ej}
\varepsilon_j(T,\mu_B) \ = \  \frac{d_j}{2\pi^2} \int_0^{\infty}k^2dk 
\sqrt{m_j^2+k^2}
\left[\exp\left(\frac{\sqrt{m_j^2+k^2}-\mu_j}{T}\right) \pm
 1 \right]^{-1}.
\end{equation}
The number of hadrons $N_{h}(T,\mu_B)$ is given by
\begin{equation}
\label{Nh}
N_{h}(T,\mu_B) = V \sum_{j} n_j(T,\mu_j).
\end{equation}
The sum in equations (\ref{Eth}) and (\ref{Nh}) runs over all
known stable hadron and resonance species \cite{pdg}.

Equation (\ref{f-o_crit}) defines the curve in the $(T,\mu_B)$
plane along which the freeze-out parameters fluctuate. To
compare our model with the experiment the relation of each
point on this curve to the transverse energy $E_T$ of neutral
hadrons at given impact parameter $b$ is needed.
The point $(\overline{T},\overline{\mu}_B)$ corresponds
to the average value $\overline{E}_T$, which in the framework
of wounded nucleon model \cite{Bialas} is proportional to the
number of participant nucleons:
\begin{equation}
\label{E_Tav}
\overline{E}_T = q N_p(b).
\end{equation}
The fluctuations of $E_T$ are known to be Gaussian-distributed
\cite{Kharzeev}:
\begin{equation}
\label{PMB}
P_{MB}(E_T|b) =
\frac{1}{\sqrt{2 \pi q^2 a N_p(b)}} \exp \left( - \  \frac{[E_T -
q N_p(b)]^2}{2 q^2 a N_p(b)} \right).
\end{equation}
The subscript `MB' states for `minimum bias'. This probability
distribution is valid for the event sample collected with the
`minimum bias' trigger, which includes all\footnote{In contrast,
the dimuon trigger requires production of a dimuon pair.} events
having the given value of $E_T$. The parameter values $q=0.274$
GeV and $a=1.27$ \cite{Chaurand} are fixed from the minimum bias
transverse energy distribution \cite{evidence}. The deviations of
$E_T$ from its central value are mostly due to change of the
number of neutral hadrons rather than due to change of average
energy per hadron. Therefore, $E_T$ is approximately proportional
to the total multiplicity of final hadrons: $E_T \sim
N_{h}^{tot}(T,\mu_B)$. Due to resonance decay,
$N_{h}^{tot}(T,\mu_B)$ is larger than the number of particles in
the hadron gas at freeze-out $N_{h}(T,\mu_B)$. Namely, every term
in the sum (\ref{Nh}) should be multiplied by a factor $g_j \ge
1$:
\begin{equation}
g_j= \left\{
\begin{array}{ll}
1 & \mbox{for stable and weakly decaying
particles;} \\
\sum_{k} k R(j  \rightarrow k h) & \mbox{for resonances.}
\end{array}
\right.
\end{equation}
Here $R(j  \rightarrow k h)$ is the probability that the resonance
$j$ produces $k$ hadrons in the final state.

Hence, the average number  of $J/\psi$ mesons,
$\langle J/\psi \rangle_{AB(b)}^{(E_T)}$,
produced in an $A+B$ collision at fixed
impact parameter $b$ and transverse energy $E_T$ can be found in the
following way:
\begin{enumerate}
\item
$T$ and $\mu_B$ are found as a solution of coupled transcendental
equations
\begin{eqnarray}
E_{th}(T,\mu_B)/N_{h}(T,\mu_B) &=&
E_{th}(\overline{T},\overline{\mu}_B)/
N_{h}(\overline{T},\overline{\mu}_B) \label{f-o} \ ,\\
N_{h}^{tot}(T,\mu_B)/ N_{h}^{tot}(\overline{T},\overline{\mu}_B)
&=& E_T/\overline{E}_T \label{xi2} \  .
\end{eqnarray}
The first of the above equations expresses the freeze-out criterion
(\ref{f-o_crit}): constant thermal energy per hadron. The second
one appears due to the proportionality
between the multiplicity of final particles and the measured neutral
transverse energy.
\item The values of $T$ and $\mu_B$ found from (\ref{f-o})
and (\ref{xi2}) are substituted into the formula (\ref{Jpsi}) giving the desired
value of $\langle J/\psi \rangle_{AB(b)}^{(E_T)}$.
\end{enumerate}

In real experiment, the impact parameter is not fixed.
The relevant quantity, the differential cross section of the $J/\psi$
production in collisions of nuclei A and B with respect to the transverse energy
$E_T$, is obtained by integrating
$\langle J/\psi \rangle_{AB(b)}^{(E_T)}$ over the impact parameter $b$:
\begin{equation}
\label{sigma_Jpsi} \frac{d \sigma_{AB \rightarrow J/\psi}}{d E_T}
= 2 \pi \int_{0}^{\infty} d b b \  \langle J/\psi
\rangle_{AB(b)}^{(E_T)} \  P_{J/\psi}(E_T|b).
\end{equation}
Here $P_{J/\psi}(E_T|b)$ is the probability to produce
neutral transverse energy $E_T$ at fixed impact parameter $b$, provided
that a $J/\psi$ particle is produced in the same event.

Formula (\ref{sigma_Jpsi}) refers to the {\it total} number of
produced $J/\psi$'s, while the measured quantity is the number
of $\mu^+\mu^-$ pairs originating from their decay and satisfying
certain  kinematics conditions. Therefore, the cross-section
(\ref{sigma_Jpsi}) should be
multiplied by the probability of $J/\psi$ decay into a dimuon pair
$B^{J/\psi}_{\mu\mu}= (5.88 \pm 0.10) \%$ \cite{pdg} and by a
factor $\eta$, the fraction of the dimuons satisfying the kinematical
conditions of the NA50 spectrometer. Because the considered $J/\psi$
production mechanism is completely different from the ``standard''
production in hard collisions, the rapidity distribution
of $J/\psi$'s may be very different from that in N+N collisions.
Consequently, $\eta$ should not be necessarily
equal to $\eta_{NN} \approx
0.24$ which is obtained from Schuler's parameterization \cite{Schuler}
of proton-proton data.
Therefore, $\eta$ will be treated as
one more free parameter.

Finally, to make a comparison with the NA50 data, the $J/\psi$
cross section should be divided by the Drell-Yan cross section.
Similarly to (\ref{sigma_Jpsi}), it is found from
\begin{equation}
\label{sigma_DY} \frac{d \sigma_{AB \rightarrow DY'}}{d E_T} =
\int d^2 b  \ \langle DY' \rangle_{AB(b)} \ P_{DY}(E_T|b).
\end{equation}
Here quantity $P_{DY}(E_T|b)$ is similar to $P_{J/\psi}(E_T|b)$,
$\langle DY' \rangle_{AB(b)}$ is the average number of Drell-Yan
pairs produced in an A+B collision at fixed impact parameter $b$.
The prime means that only the pairs satisfying the kinematical
conditions of the NA50 spectrometer are taken into account.
Similarly to $c\bar{c}$ pairs, the number of Drell-Yan pairs is
proportional to the number of primary nucleon-nucleon collisions:
\begin{equation}
\langle DY' \rangle_{AB(b)} = A B \sigma^{NN}_{DY'} T_{AB}(b),
\end{equation}
where $\sigma^{NN}_{DY'}$ is the production cross section of Drell-Yan pairs
in nucleon-nucleon collisions. This quantity is isospin dependent, therefore
the average value should be used:
\begin{equation}
\sigma^{NN}_{DY'}=\frac{\sigma^{AB}_{DY'}}{AB}.
\end{equation}
For the case of Pb+Pb collisions,
$A=B=208$ and
$\sigma^{PbPb}_{DY'} = 1.49 \pm 0.13 $ $\mu$b \cite{xsections}.

Hence, the following formula will be used to fit the NA50 data:
\begin{equation}
\label{Rb}
R(E_T) = \eta B^{J/\psi}_{\mu\mu}
\left( \frac{d \sigma_{AB \rightarrow J/\psi}}{d E_T} \right) \left/
\left( \frac{d \sigma_{AB \rightarrow DY'}} {d E_T} \right) \right.
 \ .
\end{equation}

\begin{figure}[p]
\begin{center}
%\vfill
%\leavevmode
\epsfig{file=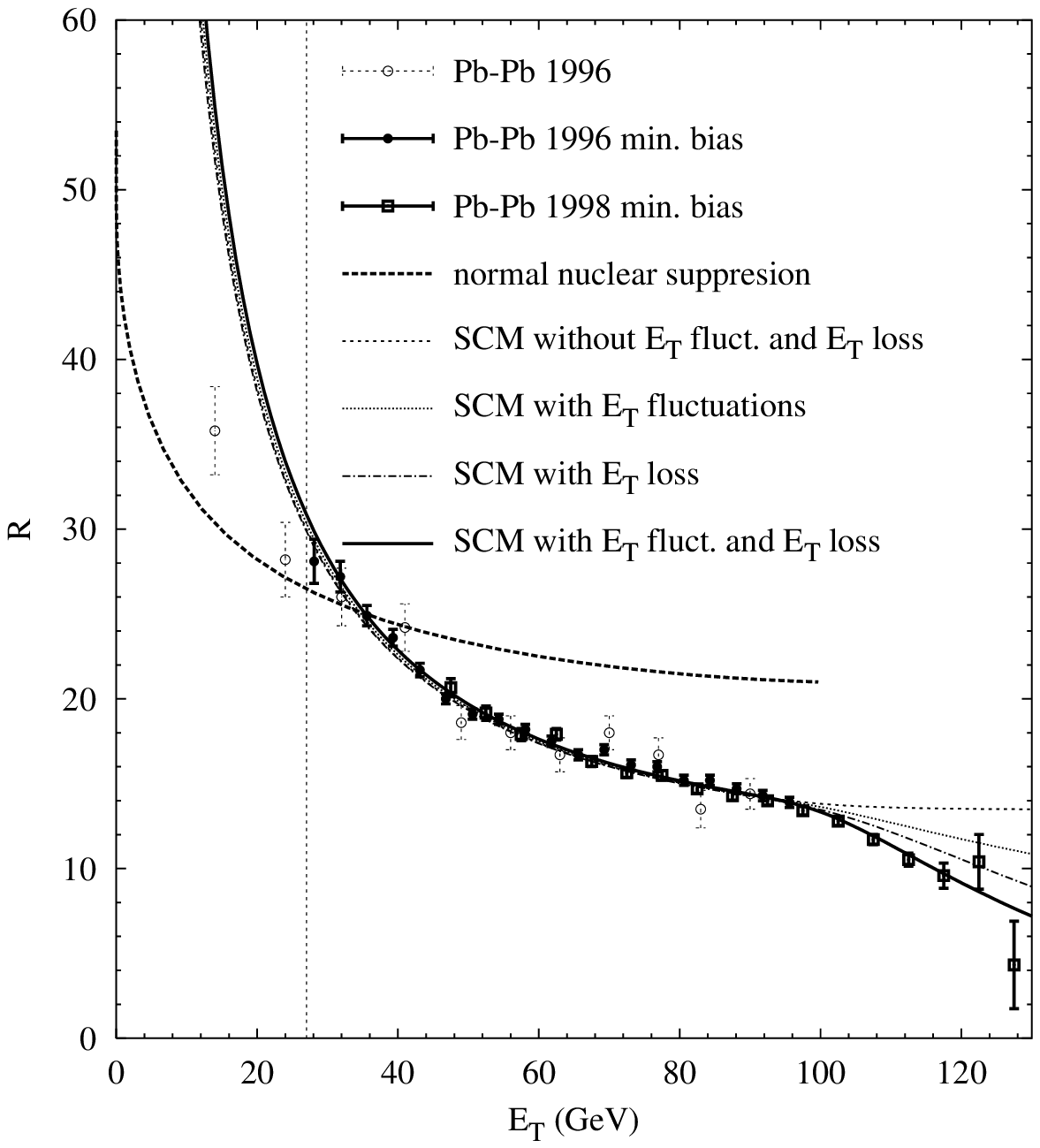,height=17cm}
\mbox{}\\
\vfill 
\mbox{}\\[1cm]
\caption{The dependence of the $J/\psi$ over Drell-Yan 
ratio $R$ on the transverse energy. The points with error bars are 
the NA50 data. 
%{\cite{anomalous,threshold,evidence}. 
The lines correspond 
to different versions of the statistical $c\bar{c}$ coalescence model
(see text for details) and to the normal nuclear suppression mechanism.  
The normal nuclear suppression curve is
obtained at $\sigma_{abs} = 6.4$~mb, where $\sigma_{abs}$ is the absorption
cross section of preresonant charmonia by sweeping nucleons.
The vertical line shows the applicability domain of the statistical 
$c\bar{c}$ coalescence model  $N_p > 100$. \label{f1} }
\end{center}
\end{figure}

As the first step, we assume that both $P_{J/\psi}(E_T|b)$ and
$P_{DY}(E_T|b)$ in (\ref{sigma_Jpsi}) and (\ref{sigma_DY})
coincide with the 'minimum bias' probability (\ref{PMB}). The
result of the calculations is shown in Fig. \ref{f1} with the
dotted line. Calculations with $T \equiv \overline{T}$ and $\mu_B
\equiv \overline{\mu}_B$, i.e. ignoring the fluctuations, were
done for comparison (the thin dashed line). As is seen, the
influence of fluctuations becomes essential only at $E_T > 100$.
This can be explained as follows. If $E_T$ is not very large, $E_T
<< q N_p(0)$,  the integration in (\ref{sigma_Jpsi}) runs over the
region where $E_T/N_p > q$, as well as over the one with $E_T/N_p
< q$. The central point of the Gaussian (\ref{PMB}), $E_T/N_p =
q$, corresponds to $T = \overline{T}$ and $\mu_B =
\overline{\mu}_B$. Deviations of $T$ and $\mu_B$ in both
directions from their central values almost cancel each other.
Therefore, the result does not differ essentially from that
obtained at $T \equiv \overline{T}$ and $\mu_B \equiv
\overline{\mu}_B$.

In contrast, when $E_T$ approaches the value $q N_p(0)$ or exceeds
it, the cancellation does not take place any more. At large $E_T$,
the region $E_T/N_p \lesssim q$ falls into nonphysical domain
$N_p > N_p(0)$. Only the tail of the Gaussian
(\ref{PMB}) with $E_T/N_p > q$ contributes to the integral
(\ref{sigma_Jpsi}). This corresponds to fluctuational tail with $T
> \overline{T}$ and $\mu_B < \overline{\mu}_B$.

Although both the thermal 'density' of $J/\psi$,
$n_{J/\psi}^{tot}(T,\mu_B)$, as well as that of the open charm,
$n_O(T,\mu_B)$, are very sensitive to the temperature, their ratio
in the formula (\ref{Jpsi}),
\begin{equation}
\frac{N_{J/\psi}^{tot}(T,\mu_B)}{(N_O(T,\mu_B)/2)^2} =
\frac{1}{V}
\frac{n_{J/\psi}^{tot}(T,\mu_B)}{(n_O(T,\mu_B)/2)^2},
\end{equation}
is much less sensitive. The additional $J/\psi$ suppression takes
place mostly due to increasing of the system volume $V$. Smaller
$\mu_B$ leads to smaller baryon density $n_B(T,\mu_B$) in
(\ref{bnc}). At fixed $N_p$ this results in a larger volume.
Decreasing the number of $J/\psi$ with increasing volume at fixed
number of $c\bar{c}$ pairs is intuitively clear: the larger is the
volume the smaller is the probability that $c$ and $\bar{c}$ meet
each other and form a quarkonium state.

As is seen from figure \ref{f1}, the suppression due to the $T$
and $\mu_B$ fluctuations is not strong enough to explain the
experimental data at $E_T > 100$ GeV. The reason is that we have
assumed that $P_{J/\psi}(E_T|b) \equiv P_{MB}(E_T|b)$. In fact,
however, production of a $J/\psi$ particle costs a fraction of
energy of the colliding nuclei. Therefore, the multiplicity of
ordinary hadrons (and consequently $E_T$) in a $J/\psi$ event is,
in average, smaller than that in a 'minimum bias' event (i.e.
without registering $J/\psi$)\footnote{The 'minimum bias'
probability distribution $P_{MB}(E_T|b)$ is  basically related to
the events without $J/\psi$ or a Drell-Yan pair in the final
state. Only a tiny fraction of Pb+Pb collisions results in
production of charmonia or Drell-Yan pairs, therefore their
influence on the shape of the minimum bias probability
distribution is negligible.} at the same impact parameter. Hence,
as was pointed out in \cite{E_T_loss}, the probability
$P_{J/\psi}(E_T|b)$ is slightly different from $P_{MB}(E_T|b)$.
Similarly to \cite{E_T_loss}, we assume that the nucleon pair
that have produced a $J/\psi$ meson does not contribute to the
transverse energy. Therefore, the probability $P_{J/\psi}(E_T|b)$
is obtained from formula (\ref{PMB}) by the replacement
$N_p(b) \rightarrow N_p(b)-2$. This induces a tiny (about $0.5$\%)
shift of the $E_T$ probability distribution.

The transverse energy distribution of the Drell-Yan event sample
$P_{DY}(E_T|b)$ should also differ from $P_{MB}(E_T|b)$ because of
the energy loss for the Drell-Yan pair production.
However, in the large $E_T$ region, where the
mentioned difference is important, our result is to be compared with the
data obtained from the {\it minimum bias analysis}. This means that
the so called "theoretical Drell-Yan" instead of the physical one
was used as a reference for $J/\psi$ suppression pattern.
This "theoretical Drell-Yan" was obtained neglecting the $E_T$ loss in the
dimuon data sample. Therefore, to be consistent with the data presented
by the NA50, we put $P_{DY}(E_T|b) \equiv P_{MB}(E_T|b)$.

The result of the calculations taking into account the modification
of $P_{J/\psi}(E_T|b)$ (ignoring, for the moment, fluctuations of
$T$ and $\mu_B$) is shown in figure \ref{f1} with the dash-dotted line.
Again, a  sizable effect (compare with the thin dashed line) is
observed only at $E_T>100$. This is explained by the following
properties of the Gaussian distribution. Let us consider a ratio of 
two Gaussians. One of them is shifted with respect to another one:
\begin{equation}
\frac{\exp \left\{  - \left[ x - (x_0 - \Delta)\right]^2 \right \} }
{\exp \left\{ -\left[ x - x_0 \right]^2  \right \}} =
\exp \left\{ -2 \Delta \left[  x - (x_0 - \Delta) \right]  \right \}
\label{rgauss}.
\end{equation}
If the shift $\Delta$ is small, the ratio (\ref{rgauss})
is very close to $1$
near the central point $x_0$. But, due to exponential behavior,
it differs essentially
from unity at large distances from $x_0$, which
corresponds to tails of the Gaussians. As was mentioned above, the
shift of $P_{J/\psi}(E_T|b)$ with respect to $P_{MB}(E_T|b)$ is
about $0.5$\%. Therefore, when the main contribution to the integral 
(\ref{sigma_Jpsi}) comes from the vicinity of the central value of 
the Gaussian (\ref{PMB}), the effect is small.
In contrast, at large $E_T$, when the integration runs only over
the tail of the Gaussian, the effect becomes essential.

Taking into account both effects, $T$ and $\mu_B$ fluctuations and $E_T$
loss in the $J/\psi$ event sample, one finds excellent agreement with
the fitted data including high $E_T$ ($E_T > 100$ GeV)
region  (the thick solid line in figure \ref{f1}).
The free parameters $\sigma^{NN}_{c\bar{c}}$ and $\eta$ were fixed
by fitting the model to the NA50 data\footnote{Only the
complete (i.e. including  $T$ and $\mu_B$ fluctuations and $E_T$
loss) model has been fitted to the data. The calculations
shown by the thin dashed, dotted and dash-dotted
lines in figure \ref{f1} has been done with the parameter set fixed
from the complete model fit.}.

As was mentioned above, the statistical coalescence model is not
expected to describe small systems. This can be seen from the $\psi'$
data \cite{Br1}. In the framework of this model
the multiplicity of $\psi'$ is given by formula (\ref{Jpsi})
with the replacement $N_{J/\psi}^{tot} \rightarrow N_{\psi'}$.
Consequently, the $\psi'$ to $J/\psi$ ratio as a function of the
centrality should be constant and equal to its thermal equilibrium
value. The experimental data \cite{Jaipur} (see also a compilation
in  \cite{Br1}) are consistent with this picture only at
rather large numbers of participants  \cite{Shuryak}. Therefore
our fit is restricted to the region
\begin{equation}
\label{apldom}
E_T > 27 \mbox{ GeV,}
\end{equation}
which corresponds to $N_p \agt 100$. Note that all NA50 data
except the two leftmost points from the 1996 standard analysis set
lie in the region (\ref{apldom}).

To check the robustness of our model with
respect to possible uncertainties we use two independent freeze-out
parameter sets. The first one \cite{Becattini} has been obtained
assuming strangeness and antistrangeness
suppression by a factor $\gamma_s$:
\begin{equation}
\label{set_Bec} \overline{T} = 158 \ {\rm MeV}, \ \ \ \
\overline{\mu}_B = 238 \ {\rm MeV} \ \ \ \ \gamma_s=0.79.
\end{equation}
In the second one \cite{Br2} the complete strangeness equilibration
has been supposed:
\begin{equation}
\label{set_Br}
\overline{T} = 168 \ {\rm MeV}, \ \ \ \
\overline{\mu}_B = 266 \ {\rm MeV} \ \ \ \
\gamma_s \equiv 1.
\end{equation}
The predicted effective $c\bar{c}$ production cross section as
well as the fit quality appeared to be insensitive to the
freeze-out parameters:
\begin{equation}
\sigma^{NN}_{c\bar{c}} =  35.7
\begin{array}{l}
{ \scriptstyle  + 10.8 } \\[-1.5ex]
{ \scriptstyle  - \  8.8  }
\end{array}
\ \mu\mbox{b}
\ , \ \ \
\chi^2/\mbox{dof} = 1.06
\end{equation}
for both sets (\ref{set_Bec}) and (\ref{set_Br}).
The parameter $\eta$ only slightly depends on
$\overline{T}$ and $\overline{\mu}_B$:
\begin{equation}
\eta =  0.134
\begin{array}{l}
{ \scriptstyle  + 0.061 } \\[-1.5ex]
{ \scriptstyle  - 0.042 }
\end{array}
\end{equation}
for the set (\ref{set_Bec}), and
\begin{equation}
\eta =  0.126
\begin{array}{l}
{ \scriptstyle + 0.059} \\[-1.5ex]
{ \scriptstyle - 0.039}
\end{array}
\end{equation}
for the set (\ref{set_Br}).

As it was already pointed out in our previous papers
\cite{Go:00,Ko:01}, the model predicts a rather large enhancement of
the total charm
(by  a factor of about $3.5$ within the rapidity
window of the NA50 spectrometer). A direct measurement of the open
charm in Pb+Pb collisions at the CERN SPS would allow for a test of 
this prediction.

Extrapolation of the model to low $E_T$ region reveals strong
disagreement with the experimental data (about 3.5 standard
deviations at the leftmost experimental point). This suggests that
the charmonium production mechanism in central and peripheral
Pb+Pb collisions may be of a very different nature. The `standard'
mechanism, formation of pre-resonant charmonium states in primary
nucleon collisions and their subsequent suppression in nuclear
medium, seems to dominate peripheral collisions. In contrast, 
the present statistical $c\bar{c}$ coalescence model assumes the 
formation of a deconfined quark-gluon medium in central collisions.
This medium destroys {\it all} \footnote{A different possibility has been
considered in \cite{Rapp}. It has been assumed that both direct and 
statistical coalescence mechanisms of $J/\psi$ production are present at all 
energies and at all collision centralities. No open charm enhancement has been 
permitted in \cite{Rapp}. Therefore the contribution of statistical 
coalescence mechanism has been found to be small at the SPS. Then, the 
primordial  $J/\psi$'s that survived 
suppression in the quark-gluon plasma dominate  even in the most 
central Pb+Pb collisions.  
Such an approach is, however, hardly able to describe the
centrality dependence of the $J/\psi$ suppression pattern.}
primary $c\bar{c}$ bound
states. The final state charmonia are formed at a later stage of the 
reaction from $c$ and $\bar{c}$ quarks available in the medium. The
distribution of $c$ and $\bar{c}$ over charmonia and open charm
hadrons follows from the laws of statistical mechanics.

In conclusion, we have shown that the statistical $c\bar{c}$
coalescence model
provides a quantitative description of the
`anomalous $J/\psi$ suppression' in Pb+Pb collisions at CERN SPS.
In particular, the `second drop' of the $J/\psi$ suppression
pattern, at $E_T\approx 100$~GeV,
can be naturally explained within the model  as the result of
two simultaneous effects: the fluctuations of the freeze-out
thermodynamic parameters and transverse energy losses in the dimuon
events with respect to the minimum bias ones.  
The model predicts a rather strong enhancement of the open charm. Therefore
a direct measurement of the
open charm would be a crucial test of 
the statistical $c\bar{c}$ coalescence model  
at the SPS.

\acknowledgments
The authors
are thankful to K.A.~Bugaev, P.~Bordalo,
M. Ga\'zdzicki and L.~McLerran  for comments and
discussions.
We acknowledge the financial support of
the Alexander von Humboldt Foundation, DFG, GSI and BMBF, Germany.
The research described in this publication was made possible in part
by Award \# UP1-2119 of the U.S. Civilian Research and Development
Foundation for the Independent States of the Former Soviet Union
(CRDF) and INTAS grant 00-00366 and INTAS grant 00-00366.

\nopagebreak

\widetext

\end{document}